\documentclass[a4paper,11pt]{article}
\pdfoutput=1 % if your are submitting a pdflatex (i.e. if you have
\usepackage{jheppub} % for details on the use of the package, please
\usepackage{multirow}
\usepackage{nicefrac}
\newcommand{\ba}{\begin{eqnarray}}
\newcommand{\ea}{\end{eqnarray}}

\usepackage[T1]{fontenc} % if needed

\title{\boldmath On preheating after inflation in scalar-tensor theories of gravity}

\author[a]{Peeravit Koad,}
\author[b]{Jatechan Channuie,}
\author[b,c]{Phongpichit Channuie}

\affiliation[a]{School of Informatics, Walailak University, Thasala, Nakhon Si Thammarat, \\80160, Thailand}
\affiliation[b]{School of Science, Walailak University, Thasala, Nakhon Si Thammarat, \\80160, Thailand}
\affiliation[c]{College of Graduate Studies, Walailak University, Thasala, Nakhon Si Thammarat, \\80160, Thailand}

\emailAdd{harrykoad@gmail.com, jatechan.c@gmail.com, channuie@gmail.com}

\abstract{In this work, we investigate the preheating mechanism after inflation in general scalar-tensor theories of gravity. In the present scenario, we consider two-field scenario and then derive the potential of exponential and hyperbolic tangent forms. We study the evolution of the background system when the back reaction on the background field is neglected. We examine the particle production due to parametric resonances in both models. We find that in Minkowski space the stage of parametric resonances can be described by the Mathieu equation. Finally, we demonstrate that parametric resonances in our models are sufficiently broad possible for the exponential growth of the number of particles. We also discuss the case in which the expansion of space is taken into account.}


\begin{document} 
\maketitle
\flushbottom

%%%%%%%%%%%%%%%%%%%%%%%
\section{Introduction}
%%%%%%%%%%%%%%%%%%%%%%%
In the standard framework of cold inflation, the universe will pass through the period of reheating of which the inflaton field decays into elementary particles populating the Universe. The instructive idea of mechanism for reheating was proposed, for instance, by the author of \cite{Linde:2005ht} in which reheating occurs due to particle production by the oscillating scalar field. However, in various inflationary models, the first stages of reheating occur in
a regime of a broad parametric resonance. This was understood after a proposal of Ref.\cite{Kofman:1997yn,Shtanov:1994ce,Kofman:1994rk,Traschen:1990sw}. During this preceding evolutionary phase so-called a preheating stage, particles are explosively produced due to the parametric resonance. Likewise, the energy transfer from the inflaton field to other particles during preheating is extremely efficient.

Regarding existing literature, there were many analytical works which examined the preheating mechanism, e.g. \cite{Greene:1997fu,Kaiser:1995fb,Son:1996uv}. The properties of resonance with non-minimally coupled scalar field ${\chi}$ in preheating phase have been carried out by the authors of \cite{Tsujikawa:1999jh}. Here the effective resonance is possible only by inclusion of a non-minimal coupling $\xi R\chi^{2}$ term with a sizeable range of parameter $\xi$. Higher-curvature inflation models with $(R+\alpha^{n}R^{n})$ allowing to study a parametric preheating of a scalar field coupled non-minimally to a spacetime curvature were also investigated \cite{Tsujikawa:1999iv}. In Ref.\cite{vandeBruck:2016leo}, the authors studied preheating effects in the extended Starobinsky model with an additional scalar field which interacts directly with the inflaton field via a four-leg interaction term.

Preheating after Higgs inflation with self-resonance and gauge boson production has been carried out in Ref.\cite{Sfakianakis:2018lzf}. Another interesting paradigm was proposed by \cite{Bezrukov:2008ut,Garcia-Bellido:2008ycs}. In this scenario, they studied preheating mechanism of which the standard model Higgs, strongly non-minimally coupled to gravity, plays the role of the inflaton. Consequently, they discovered that the universe does reheat through which perturbative and non-perturbative effects are mixed. Additionally, based on Higg inflation, preheating effects gained much interest, see for example \cite{Hamada:2020kuy,Rubio:2019ypq}. Moreover, the authors of \cite{DeFelice:2012wy} have investigated the production of particles due to parametric resonances in 3-form field inflation and found that this process is more efficient compared to the result of the standard-scalar-field inflationary scenario, e.g. \cite{Kofman:1997yn,Shtanov:1994ce,Kofman:1994rk}, in which the broad resonance tends to disappear more quickly. Interestingly, regarding multi-field inflation, preheating mechanism in asymmetric $\alpha$-attractors has also been discussed \cite{Iarygina:2020dwe}, see also Ref.\cite{Nguyen:2019kbm} for preheating after multifield inflation; while Palatini formalism of gravitational dark matter production during preheating stage is worth mentioning \cite{Karam:2020rpa}. The study of preheating in the Palatini formalism with a quadratic inflaton potential and an added $\alpha\,R^2$ term has been discussed in Ref.\cite{Karam:2021sno}.

In our work, the preheating mechanism after inflation in general scalar-tensor theories of gravity is investigated. This paper is organized as follows. In Sec.(\ref{s2}), we briefly review a two-field inflationary scenario and then derive the potential of exponential and hyperbolic tangent forms. In Sec.(\ref{s3}), we employ an analytical approach to study the preheating for model of inflation in general scalar-tensor theories of gravity and study parametric resonances of models when a inflaton field $\phi$ coupled to another scalar field $\chi$ with the interaction term $g^{2}\phi^{2}\chi^{2}$. In Sec.\ref{ch4}, we consider the case in which the expansion of space is taken into account. Finally, we give our findings in the last section.

%%%%%%%%%%%%%%%%%%%%%%%%%%%%%%%
\section{Formalism}\label{s2}
%%%%%%%%%%%%%%%%%%%%%%%%%%%%%%%
In this work, we study the preheating process after inflation in general scalar-tensor theories of gravity. In this section, we will follow the work proposed by Ref.\cite{Kaiser:2010ps} and focus on the single field inflationary scenario in which the only inflaton is nonminimally coupled to gravity, while another scalar field is produced by the oscillation of the inflaton field. Noting that a simplified two-field model has been studied often in the literature, e.g., see \cite{Bassett:1997az,Tsujikawa:2002nf,Amin:2014eta}. Moreover, the preheating effect after multifield inflation has been so far studied, see Ref.\cite{DeCross:2015uza}. In the following, we choose a general form of the 4D action of the two-field system in the Jordan (J) frame:
\begin{eqnarray}
\mathcal{S}_{\rm J}=\int d^{4}x \sqrt{-g}&&\Big[- f(\Phi^{i})R+ \frac{1}{2}\omega(\Phi^{i})_{ij}g^{\mu\nu}\nabla_{\mu}\Phi^{i}\nabla_{\nu}\Phi^{j} -V_{J}(\Phi^{i}) \Big]\,, \label{actionset}
\end{eqnarray}
where $\Phi^{i}=(\phi,\chi)$ with $\chi$ an additional scalar field, and $V_{J}(\Phi^{i})$ is the Jordan-frame potential with $i=1,2$. The function $f(\Phi^{i})$ and $\omega(\Phi^{i})_{ij}$ are an arbitrary function on the scalar field $\Phi^{i}$. Let us consider a typical form of $f(\Phi^{i})$ in our case in which the non-minimal couplings take the form. 
\begin{eqnarray}
f(\Phi^{i}) = \frac{1}{2}\left(M^{2}_{0} + \xi_{\Phi^{i}}(\Phi^{i})^{2}\right)\,, \label{fP}
\end{eqnarray}
where $M_{0}$ in this work is assigned to be the Planck constant $M_{p}$ and the coupling strengths $\xi_{\Phi^{i}}$ are the couplings between curvature and matter fields. In order to bring the gravitational portion of the action into the canonical Einstein-Hilbert form, we perform a conformal transformation by rescaling $\tilde{g}_{\mu\nu} = \Omega^{2}(x)g_{\mu\nu}$. Here we can relate the conformal factor $\Omega^{2}(x)$ to the nonminimal-coupling sector via
\begin{eqnarray}
\Omega^{2}(\Phi^{i}) = \frac{2}{M^{2}_{p}}f(\Phi^{i}(x))\,, \label{con}
\end{eqnarray}
By applying the conformal transformation given above, we can eliminate the nonminimal-coupling sector and obtain the resulting action in the Einstein frame \cite{Kaiser:2010ps}
\begin{eqnarray}
\mathcal{S}_{\rm E}=\int d^{4}x \sqrt{-g}&&\Big[- \frac{M^{2}_{p}}{2}{\hat R} + \frac{1}{2}{\cal G}_{ij}g^{\mu\nu}\nabla_{\mu}\Phi^{i}\nabla_{\nu}\Phi^{j} -{\cal U}(\Phi^{i})\Big]\,, \label{ef}
\end{eqnarray}
with ${\cal U}(\Phi^{i})\equiv V_{J}(\Phi^{i})/\Omega^{4}$ and ${\hat R}$ is the Ricci curvature scalar in the Einstein frame. Here we have dropped the tildes for convenience and ${\cal G}_{ij}$ is given by
\begin{eqnarray}
{\cal G}_{ij} = \frac{M^{2}_{\rm P}}{2f}{\omega(\Phi^{i})_{ij}} + \frac{3}{2}\frac{M^{2}_{\rm P}}{f^{2}}f_{,i}f_{,j}\,, \label{Gij}
\end{eqnarray}
where $f_{,i}=\partial f/\partial\Phi^{i}$. In our analysis, it is more convenient to express ${\cal G}_{ij}$ in terms of $\Omega$. Substituting Eq.(\ref{con}) into Eq.(\ref{Gij}), we come up with the following relation:
\begin{eqnarray}
{\cal G}_{ij} = \bigg[\frac{{\omega_{ij}}}{\Omega^{2}}+ 6M^{2}_{p}\frac{\Omega_{,i}\Omega_{,j}}{\Omega^{2}} \bigg]\,, \label{Gijo}
\end{eqnarray}
where an argument of $\omega$ and $\Omega$ is understood. In our case, the above quantity can be explicitly recast in terms of the fields $(\phi,\chi)$ as
\begin{eqnarray}
{\cal G}_{\phi\phi} &=& \bigg[\frac{\omega(\phi)}{\Omega^{2}}+ 6M^{2}_{p}\frac{\Omega_{,\phi}\Omega_{,\phi}}{\Omega^{2}} \bigg]\,, \label{pp}\\
{\cal G}_{\phi\chi} &=& \bigg[ 6M^{2}_{p}\frac{\Omega_{,\phi}\Omega_{,\chi}}{\Omega^{2}} \bigg]={\cal G}_{\chi\phi}\,, \label{pc}\\
{\cal G}_{\chi\chi} &=& \bigg[\frac{\omega(\chi)}{\Omega^{2}}+ 6M^{2}_{p}\frac{\Omega_{,\chi}\Omega_{,\chi}}{\Omega^{2}} \bigg]\,.\label{cc}
\end{eqnarray}
We assume a particular scenario in which $\omega(\phi)$ solely depends on $\phi$ of the form
\begin{eqnarray}
\omega(\phi)=\frac{M^{2}_{p}}{\xi}\Omega_{,\phi}\Omega_{,\phi}\,, \label{om}
\end{eqnarray}
and take $\omega(\chi)=1$. Therefore for ${\cal G}_{\phi\phi}$ and $\Omega=\Omega(\phi)$, there exists an exact relationship between $\phi$ and $\hat{\phi}$ via:
\begin{eqnarray}
\frac{d\hat{\phi}}{d\phi}=\sqrt{\frac{\omega(\phi)}{\Omega^{2}}+ 6M^{2}_{p}\frac{\Omega_{,\phi}\Omega_{,\phi}}{\Omega^{2}}}=\sqrt{6\alpha}M_{p}\frac{\Omega_{,\phi}}{\Omega}\,,
\end{eqnarray}
where $\alpha=1+(6\xi)^{-1}$. We can imply solve the above equation to obtain
\begin{eqnarray}
\hat{\phi}=\sqrt{6\alpha}\,M_{p}\ln \Omega(\phi),\,\,\,\,\,\,\Omega(\phi)=e^{\sqrt{1/6\alpha}\,\hat{\phi}/M_{p}}\,.
\label{conso}
\end{eqnarray}
and
\begin{eqnarray}
{\cal U}=\frac{{V_{J}(\hat{\phi},\hat{\chi})}}{\Omega^{4}(\hat{\phi})}\,,
\label{conso1}
\end{eqnarray}
Since $\Omega=\Omega(\phi)$, we also find a relationship between $\chi$ and $\hat{\chi}$ by taking
\begin{eqnarray}
\frac{d\hat{\chi}}{d\chi}=\sqrt{\frac{1}{\Omega^{2}}}\quad{\rm or}\quad \chi=\Omega\,\hat{\chi}=e^{\sqrt{1/6\alpha}\,\hat{\phi}/M_{p}}\hat{\chi}\,.\label{cch}
\end{eqnarray}
From Eq.(\ref{con}), we can write an explicit form of $\Omega$ to obtain
\begin{eqnarray}
\Omega^{2}(\phi)=\frac{2}{M^{2}_{p}}f(\phi) = 1 + \frac{\xi \phi^{2}}{M^{2}_{p}}\,, \label{fPh}
\end{eqnarray}
so that we can write
\begin{eqnarray}
\phi^{2}=\frac{M^{2}_{p}}{\xi}\big(\Omega^{2}(\phi)-1\big)=\frac{M^{2}_{p}}{\xi}\bigg(e^{\sqrt{4/6\alpha}\,\hat{\phi}/M_{p}}-1\bigg)\,. \label{fPh1}
\end{eqnarray}
Therefore, the action written in terms of the fields $(\hat{\phi},\hat{\chi})$ takes the form
\begin{eqnarray}
\mathcal{S}_{\rm E}=\int d^{4}x \sqrt{-g}&&\Big[- \frac{M^{2}_{p}}{2}{\hat R} + \frac{1}{2}g^{\mu\nu}\nabla_{\mu}\hat{\phi}\nabla_{\nu}\hat{\phi} + \frac{1}{2}g^{\mu\nu}\nabla_{\mu}\hat{\chi}\nabla_{\nu}\hat{\chi} -{\cal U}(\hat{\phi},\hat{\chi})\Big]\,. \label{actionE2}
\end{eqnarray}
Notice that the field $\hat{\phi}$ is directly related to the conformal transformation factor $\Omega$. The resulting action in the Einstein frame yields the following equations of motion for $\hat{\phi}$ and $\hat{\chi}$, respectively
\begin{eqnarray}
&& \ddot{\hat{\phi}}+3\frac{\dot{a}}{a}\dot{\hat{\phi}}- \frac{1}{a^{2}}\nabla^{2}\hat{\phi}+\frac{\partial {\cal U}}{\partial\hat{\phi}} = 0 \,, \label{EoMphi}\\
&& \ddot{\hat{\chi}}+3\frac{\dot{a}}{a}\dot{\hat{\chi}} - \frac{1}{a^{2}}\nabla^{2}\hat{\chi}+\frac{\partial {\cal U}}{\partial\hat{\chi}} = 0 \,. \label{EoMchi}
\end{eqnarray}
Since we are interested in a preheating effect after inflation, we assume that the spacetime and the inflaton $\phi$ give a classical background and the scalar field $\chi$ is treated as a quantum field on that background, see Ref.\cite{Channuie:2016xmq} for analysis on preheating after inflation in which the inflaton is the lightest composite state stemming from the minimal technicolor theory.

%%%%%%%%%%%%%%%%%%%%%%%%%%%%%%%%%%
\section{Parametric resonance}\label{s3}
%%%%%%%%%%%%%%%%%%%%%%%%%%%%%%%%%
In this section, we study parametric resonances of models when an inflaton field $\phi$ coupled to another scalar field $\chi$ with the interaction term $g^{2}\phi^{2}\chi^{2}$. In the first case scenario, we will choose the potential in the Jordan frame of the form:
\begin{eqnarray}
V_{J}(\phi)=V_{0}\big(1-\Omega^{2}(\phi)\big)^{2}+\frac{1}{2}g^{2}\phi^{2}\chi^{2}+\frac{1}{2}m^{2}_{\chi}\chi^{2}\,,
\end{eqnarray}
where $V_{0}$ can be constrained given in Ref.\cite{Yuennan:2022zml}. Then we obtain the exponential form of the potential, named E-model, and it is written in the Einstein frame of the form
\begin{eqnarray}
{\cal U}(\phi,\chi)&=&V_{0}\bigg(1-e^{-\frac{2\phi}{\sqrt{6\alpha}M_{p}}}\bigg)^{2}+\frac{g^{2}M^{2}_{p}}{2\xi}e^{-\frac{2\phi}{\sqrt{6\alpha}M_{p}}}\Big(e^{\frac{2\phi}{\sqrt{6\alpha}M_{p}}}-1\Big)\chi^{2}\nonumber\\&+&\frac{1}{2}m^{2}_{\chi}e^{-\frac{2\phi}{\sqrt{6\alpha}M_{p}}}\chi^{2}\,.
\label{consoE}
\end{eqnarray}
Here we have dropped \lq\lq\,\,$\hat{}$\,\,\rq\rq\,\,for convenience. In the present investigation, we assume that the back reaction of the created particles can be neglected during the process of parametric resonance. However, the back reaction effect on the model can be worth examining and this will be intentionally left for future investigation, see Ref.\cite{Kofman:1997yn} for useful reference on the topic. 

Soon after the end of inflation, the rapid decrease of the Hubble rate reduces the amplitude of the inflaton field. The effective potential of the inflaton field following from Eq.(\ref{consoE}) can be locally approximated by a quadratic form \cite{Bezrukov:2008ut,Garcia-Bellido:2008ycs}, say ${\cal U}(\phi)\sim M^{2} \phi^{2}/2$. Therefore, the Klein-Gordon equation (\ref{EoMphi}) for
the inflaton field is approximately given by
\begin{eqnarray}
\ddot{\phi}+3H\dot{\phi}+M^{2}\phi = 0\quad{\rm with}\quad M^{2}=\frac{4V_{0}}{3\alpha M^{2}_{p}}\,,
\label{Einf}
\end{eqnarray}
where $V_{0}$ is constrained to obtain \cite{Yuennan:2022zml}
\begin{eqnarray}
V_{0}\simeq \frac{4.35\times 10^{-7} \alpha M_p^4}{N^2 \left(1-\frac{0.75 \alpha }{N}\right)^4}\,.
\end{eqnarray}
In order to obtain the solution of the above
equation, we assume a power-law evolution of a scale factor $a\sim t^{p}$. This cosmological power law model is an interesting proposal for the evolution of the scalar factor, and it has been motivated by the existence of the flatness and horizon problems in standard cosmology \cite{Rani:2014sia}. In this cosmological model, it is possible to assume the following form for the evolution of the scale factor, $a (t) = a_{0}t^p$.  where $a_0$ is the present-day value of $a(t)$. It is also important to only consider $p > 0$, and this will produce an accelerating universe \cite{Rani:2014sia}. It may be noted that for  $p > 1$, the power-law cosmology can solve the horizon problem, the flatness problem, and the problem associated with age of the early universe \cite{Mannheim:1989jh,Allen:1998vx}. Power-law cosmology has been also used for analyzing the cosmological behavior in modified theories of gravity \cite{Bamba:2014mya,Bamba:2013fha}. Then an equation of motion becomes
\begin{eqnarray}
t^{2}\ddot{\phi}+3pt\dot{\phi}+t^{2} M^{2}\phi = 0\,.
\label{Einftp}
\end{eqnarray}
The above ordinary differential equation is given in a standard form given that its solution is well known. On physical grounds, the general solution of the effective equation of $\phi$ can be
expressed in terms of the Bessel functions and the physical solution to this equation is then
simply given by
\begin{eqnarray}
\phi(t)\simeq A (Mt)^{-\frac{(3p-1)}{2}}J_{\frac{(3p-1)}{2}}(Mt)\,,
\label{sop}
\end{eqnarray}
where the large argument expansion of fractional Bessel functions such that $Mt\gg 1$ is applied. Here a constant $A$ is chosen by assuming that the oscillatory behaviour starts just at the end of inflation, i.e.,
\begin{eqnarray}
\phi(t=t_{0})=\phi_{end}=\sqrt{\frac{3\alpha}{2}}M_{p}\log \left(\frac{2}{\sqrt{3\alpha}}+1\right).
\end{eqnarray}
Detailed derivation of the above result is given in Ref.\cite{Yuennan:2022zml}. For the large argument expansion, the above physical solution can be approximately represented by a cosinusoidal function \cite{Garcia-Bellido:2008ycs}. In our analysis, it takes the form
\begin{eqnarray}
\phi(t)\simeq A\,(Mt)^{\frac{-(3p)}{2}}\cos\left(M(t-t_{\rm os}) - (3p\pi/4)\right)\,,
\label{sops}
\end{eqnarray}
where a case with $p=2/3$ corresponds to a matter-dominated behavior as the universe continues to expand; while the system behaves like radiation corresponding to the case with $p=1/2$.  However, in the
present analysis, we assume the dynamics of the
oscillation by considering $p=2/3$ during the early time of the preheating phase. In this particular case, we find the physical solution of Eq.(\ref{sops}) as
\begin{eqnarray}
\phi(t)\simeq \Phi(t) \sin(M(t-t_{\rm os}))\,,
\label{sops1}
\end{eqnarray}
\begin{figure}[!h]	
	\includegraphics[width=12cm]{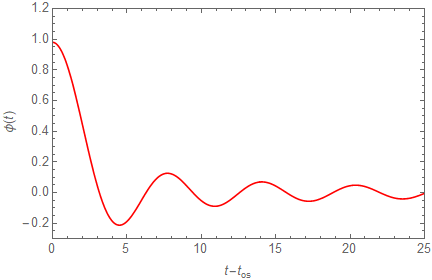}
	\centering
	\caption{We plot the approximate solution of the field $\phi(t)$ as given in Eq.(
	\ref{sops1}) using $\alpha\simeq 1.333$. The value of the scalar field here is measured in units of $M_p$ and time is measured in units of $M^{-1}$.}
	\label{plotp1}
\end{figure}
where $t_{os}$ denotes a time when the oscillating phase begins and $\Phi(t)$ is defined as
\begin{eqnarray}
\Phi(t) = \frac{\phi_{\rm end}}{M(t-t_{\rm os})}\approx \frac{\phi_{\rm end}}{2\pi \bar{N}}\,,
\label{Phi}
\end{eqnarray}
Here $\Phi(t)$ is the amplitude of oscillations, $\bar{N}$ is the number of oscillations since the end of inflation. In Fig.(\ref{plotp1}), we display the evolution of a scalar field $\phi(t)$ as described by the
approximate equation of Eq.(\ref{sops1}) with $p=2/3$. We find that the amplitude of the first oscillation drops by a factor of ten during the first oscillation.

From (\ref{EoMchi}), it is rather straightforward to derive the equation of motion for the field $\chi$ to obtain
\begin{eqnarray}
\ddot{\chi}+3H\dot{\chi} - \frac{1}{a^{2}}\nabla^{2}\chi +\Big[m^{2}_{\chi} + \sigma\phi\Big] \chi = 0\,,  \label{ST_2.2}
\end{eqnarray}
which we have defined $\sigma\equiv\sqrt{\frac{1}{6\alpha}}\frac{g^{2}M_{p}}{\xi}$. Note that in a non-flat FLRW universe, it is also possible to redefine the spatial part in terms of the new radial coordinate, the geometry will allow us to perform a similar Fourier mode decomposition using the modified plane wave functions. We then expand the scalar fields $\chi$ in terms of the Heisenberg representation to yield
\begin{eqnarray}
\chi(t,{\bf x}) \sim \int \left(a_{k}\chi_{k}(t)e^{-i{\bf k}\cdot{\bf x}} + a^{\dagger}_{k}\chi^{*}_{k}(t)e^{i{\bf k}\cdot{\bf x}}\right)d^{3}{\bf k} \,, \label{mospace}
\end{eqnarray}
where $a_{k}$ and $a^{\dagger}_{k}$ are annihilation and creation operators. We can show that $\chi_{k}$ obeys the following equation of motion:
\begin{eqnarray}
\ddot{\chi}_{k}+3H\dot{\chi}_{k} +\Big[ \frac{k^{2}}{a^{2}} + m^{2}_{\chi} + \sigma\phi\Big]\chi_{k} = 0\,. \label{Pot211}
\end{eqnarray}
Performing Fourier transformation to this equation and rescaling the field using $Y_{k} = a^{3/2}\chi_{k}$, we have
\begin{eqnarray}
\ddot{Y}_{k} + \omega^{2}_{k}Y_{k} = 0\,, \label{Pot2111}
\end{eqnarray}
where a time dependent frequency of $Y_{k}$ is given by
\begin{eqnarray}
\omega^{2}_{k} = \frac{k^{2}}{a^{2}} + m^{2}_{\chi}+ \sigma\Phi\sin(Mt)\,. \label{Pot20}
\end{eqnarray}
As is expected, Eq.(\ref{Pot2111}) describes an oscillator with a periodically changing frequency $\omega^{2}_{k} = \frac{k^{2}}{a^{2}} + m^{2}_{\chi}+ \sigma\Phi\sin(Mt)$. The physical momentum ${\bf p}$ coincides with ${\bf k}$ for Minkowski space such that $k=\sqrt{{\bf k}^{2}}$. The periodicity of Eq.(\ref{Pot2111}) may drive the parametric resonance for modes with certain
values of $k$. We will examine this behavior by
introducing a new variable, $z$ and define it via $M(t-t_{os}) =2z-\pi/2$. In the Minkowski space for which $a(t)=1$, Eq.(\ref{Pot2111}) becomes the standard Mathieu equation \cite{McLachlan1947}. It is governed by
\begin{eqnarray}
\frac{d^{2}Y_{k}}{dz^{2}} + \left(A_{k} - 2q\cos(2z)\right)Y_{k} = 0\,, \label{Mathieu1}
\end{eqnarray}
where
\begin{eqnarray}
A_{k} = \frac{4}{M^{2}}\left(k^{2} + m^{2}_{\chi}\right)\,,\,\, q = \frac{4\sigma\Phi}{M^{2}} \,, \label{Matheiu2}
\end{eqnarray}
where $M$ is defined in Eq.(\ref{Einf}).
\begin{figure}[!h]	
	\includegraphics[width=12cm]{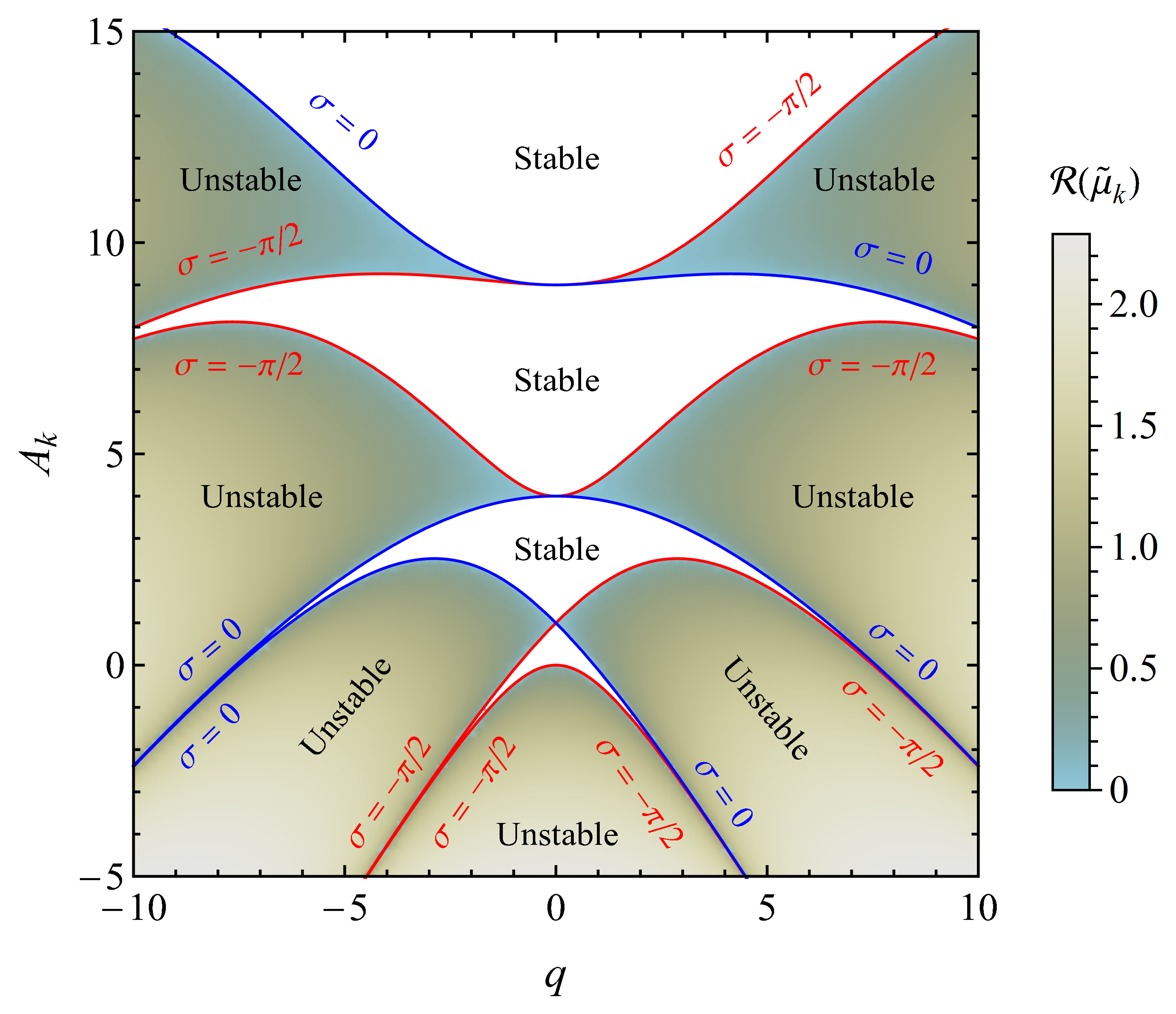}
	\centering
	\caption{We display stability-instability chart for the Mathieu equation (\ref{Mathieu1}). Empty regions are stable
while colored regions are unstable. Each instability band is spanned by $\sigma \in (-\pi/2,0)$, and the color grading indicates the coefficient of instability $|Re(\mu)|$.} \label{plotpm}
\end{figure}

The general solution of the Mathieu equation can be written in the form of $e^{\pm \mu_{k}}\,P(z)$, where $P(z+\pi) = P(z)$ \cite{McLachlan1947}. If the characteristic exponent $\mu_{k}$ has a real part, the solution of the Mathieu equation is unstable for generic initial conditions. Because $\mu_{k}$ is a function of $A_{k}$ and $q$, the instability region can be represented on the $(A_{k},\,q)$-plane, see Fig.\ref{plotpm}. In general, the strength of parametric resonance is controlled by the parameters $A_{k}$ and $q$. In order to guarantee enough efficiency for the particle production, the Mathieu equation's parameters should satisfy the broad-resonance condition, i.e. $q\gg 1$. If this is the case, a broad resonance can possibly occur for a wide range of the parameter spaces and momentum modes. Therefore, in order to satisfy a broad resonance condition, we discover that 
\begin{eqnarray}
g^{2}\gg \frac{6.07\times 10^{-7} N^2}{(\alpha -1) \log \left(\frac{1.15}{\sqrt{\alpha }}+1\right) (N-0.75 \alpha )^4}\,. \label{qq}
\end{eqnarray}
We can give an example of values of the parameters which work for our model. Taking $N=60,\,\alpha\approx 1.333$, we find $g\gg 2.80\times 10^{-5}$. Typically, in the simplest inflationary scenario including the one we are considering now, the value of the Hubble constant at the end of inflation is of the same order, but somewhat smaller, as the inflaton (effective) mass, $M$. We may also expect during this early phase of oscillation the field's kinetic energy to be roughly equal to its potential energy, and hence we can estimate the energy density of the field to be $\rho \sim M^{2}\phi^{2}\sim \frac{1}{25}{M}^{2}M^{2}_{p}$. This approximation allows us to further estimate the Hubble constant and we find that the Hubble rate would then be $H=\sqrt{\frac{1}{3 M^{2}_{p}}\rho}\sim{M}/(5\sqrt{3}) \sim 0.1 M$. Notice that the estimate for $H/M \sim 0.1$ is lower than that found in the nonminimally coupled theory in the limit $\xi_{\phi}\gg 1$, see Refs.\cite{DeCross:2015uza,Channuie:2016xmq}. For $q>0$, the solution of the Mathieu equation in the first instability band can be approximated to obtain \cite{Creminelli:2019nok}
\begin{eqnarray}
Y_{k}(z) \sim c_{+}e^{\mu_{k}z}\sin(z-\sigma)+c_{-}e^{-\mu_{k}z}\sin(z+\sigma)\,. \label{so}
\end{eqnarray}
where $\mu_{k}>0$ and $\sigma \in (-\pi/2,0)$ is a parameter, depending on $A_{k}$ and $q$. It is real inside the instability bands, as shown in the instability chart for the Mathieu equation in Fig.\ref{plotpm}. More specifically, in the first instability band one has \cite{McLachlan1947}
\begin{eqnarray}
A_{k} &=& 1 -q\cos(2\sigma) +{\cal O}(q^{2}),\\ \mu_{k} &=& -\frac{1}{2}q\sin(2\sigma)+{\cal O}(q^{2})\,, \label{Amu}
\end{eqnarray}
where the coefficients $c_{+}$ and $c_{-}$ can be determined to recover the vacuum solution at $z=0$.

\begin{figure}[t]
\begin{center}		
\includegraphics[width=0.85\linewidth]{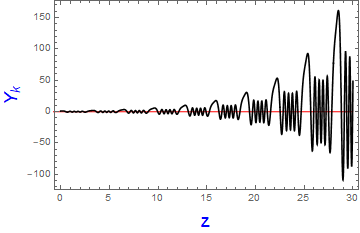}
\caption{We take $k = 4\,M\,(=m_{\chi}), q=64$, and plot the amplification of the real part of the eigenmode $Y_{k}(z)$. The exponents show the order of magnitude for each given mode of fluctuations. By comparing Fig.(\ref{Ynk}) with Fig.(\ref{plotp1}) for the equal-time interval, we see for each oscillation of the field $\phi(t)$ that the field $Y_{k}$ oscillates many times. \label{Ynk}}
\end{center}
\end{figure}

We can give an example for which the parameter $q$ takes a large value and make a plot the evolution of fluctuations $Y_{k}$. We consider the typical resonance of particle production by taking $k \sim m_{\chi}(= 4\,M)$ and $q\simeq 64$. The plot of Fig.(\ref{Ynk}) shows the amplification of the real part of the eigenmode $Y_{k}(z)$. Each peak in the amplitude of the oscillations of the field $\chi$ corresponds to a place where $\phi(t)=0$. We also see from Fig.(\ref{Ynk}) that the amplitude of the oscillation for the second mode is much larger than that of the first one; while the third one is much larger than those of the first-two modes, and so forth.  Indeed, as mentioned in Ref.\cite{Kofman:1997yn}, the effective mass of $\chi$ in our model $m^{2}_{\chi}(t) = m^{2}_{\chi}+\sigma \phi(t)$ is much greater than the inflaton mass $M$ for the main part of the period of oscillation of the field in the broad resonance regime with $q^{1/2}=\sqrt{\sigma \Phi}/M\gg 1$. Hence, the typical frequency of oscillation $\omega^{2}_{k} = \frac{k^{2}}{a^{2}} + m^{2}_{\chi}+ \sigma\phi(t)$ of the field $\chi$ is much higher than that of the field $\phi$. Within one period of oscillation of the field $\phi$ the field $\chi$ makes ${\cal O}(q^{1/2})$ oscillations. It is worth mentioning here about Fig.(\ref{Ynk}) that in standard scalar models for DE and DM there is no chaotic pattern unless the system is very sensitive to the initial values or highly parametric sensitive. A Poincare portrait of phase space as it is presented here does not show a standard pattern in the chaotic dynamical systems.

Additionally, we can expect that the growth of the modes $Y_{k}$ leads to the growth of the occupation numbers of the created particles $n_{k}(t)$. Hence, it is worth estimating the number of $\chi$-particles. Since the index $\mu_{k}$ vanishes at the edges of the resonance band and takes its maximal value of $\mu_{k}=q/2$ at $\sigma=-\pi/2$ which allows us to write
\begin{eqnarray}
\mu_{k} = \frac{q}{2}= \frac{\sqrt{3\alpha}g^{2}M^{3}_{p}}{4\sqrt{2}\xi V_{0}}\Phi\,. \label{qma}
\end{eqnarray}
In this case, the corresponding modes $Y_{k}$ grow at a maximal rate $\exp(qz/2)$, which in our case is given by $Y_{k}\sim\exp(q Mt/4)$. When the modes $Y_{k}$ grow as $\exp(qz/2)$, the number of $\chi$-particles grows as $\exp(qz)$, which in our case is equal
to 
\begin{eqnarray}
n_{k}\sim\exp(q Mt/2)=\exp\Big(\frac{\sqrt{3\alpha}g^{2}M^{3}_{p}Mt}{8\sqrt{2}\xi V_{0}}\Phi\Big)\,. \label{mu1}
\end{eqnarray}
Our basic finding is that the number of particles grows exponentially. As mentioned in Ref.\cite{Kofman:1997yn}, this process can be interpreted as a resonance with decay of two $\phi$-particles with mass $M$ to two $\chi$-particles with momenta $k=4M$.

In the second model, under the condition (\ref{om}), if we instead choose
\begin{eqnarray}
V_{J}(\phi,\chi)=V_{0}\Omega^{4}(\phi)\bigg(\frac{1-\Omega^{2}(\phi)}{1+\Omega^{2}(\phi)}\bigg)^{2}+\frac{1}{2}g^{2}\phi^{2}\chi^{2}+\frac{1}{2}m^{2}_{\chi}\chi^{2}\,,
\label{cot}
\end{eqnarray}
then we get the hyperbolic tangent form of the potential, named it as T-model, in the Einstein frame
\begin{eqnarray}
{\cal U}(\phi,\chi)&=&V_{0}\tanh^{2}\bigg(\frac{\phi}{\sqrt{6\alpha}M_{p}}\bigg)+\frac{g^{2}M^{2}_{p}}{2\xi}e^{-\frac{2\phi}{\sqrt{6\alpha}M_{p}}}\Big(e^{\frac{2\phi}{\sqrt{6\alpha}M_{p}}}-1\Big)\chi^{2}\nonumber\\&+&\frac{1}{2}m^{2}_{\chi}e^{-\frac{2\phi}{\sqrt{6\alpha}M_{p}}}\chi^{2}\,,
\label{cot2}
\end{eqnarray}
where $t_{os}$ again denotes a time when the oscillating phase begins. Soon after the end of inflation, the rapid decrease of the Hubble rate reduces the amplitude of the inflaton field. The effective potential of the inflaton field following from Eq.(\ref{cot2}) can be locally approximated by a quadratic form \cite{Bezrukov:2008ut,Garcia-Bellido:2008ycs}, say ${\cal U}(\phi)\sim {\color{blue}{{\tilde M}^{2}}} \phi^{2}/2$. Therefore, the Klein-Gordon equation (\ref{EoMphi}) for
the inflaton field of this model is approximately given by
\begin{eqnarray}
\ddot{\phi}+3H\dot{\phi}+{\tilde M}^{2}\phi = 0\quad{\rm with}\quad {\tilde M}^{2}=\frac{V_{0}}{3\alpha M^{2}_{p}}\,,
\label{Einf2}
\end{eqnarray}
where $V_{0}$ is constrained to obtain \cite{Yuennan:2022zml}
\begin{eqnarray}
V_{0}\simeq \frac{6.96\times 10^{-6}\alpha M_{p}^4}{(4N-3\alpha )^2}\,.
\end{eqnarray}
From now on, we will follow the formalism given in the preceding subsection. Hence, we can solve the ODE to obtain the physical solution of Eq.(\ref{Einf2}) to yield
\begin{eqnarray}
\phi(t)\simeq \frac{\phi_{end}}{{\tilde M}(t-t_{\rm os})} \sin({\tilde M}(t-t_{\rm os}))\,,
\label{sops2}
\end{eqnarray}
where $\phi_{end}$ of this model reads
\begin{eqnarray}
\phi(t=t_{0})=\phi_{end}\simeq \sqrt{\frac{3 \alpha }{2}} M_{p} \sinh^{-1}\left(\frac{2}{\sqrt{3\alpha}}\right)
\end{eqnarray}
Similarly to the previous case, the evolution of a scalar field $\phi(t)$ can be described by the
approximate equation of Eq.(\ref{sops2}) with $p=2/3$. We see that the amplitude of the first oscillation drops by a factor of ten during the first oscillation. Following our formalism given in the preceding subsection, the field $\chi_{k}$ obeys the following equation of motion:
\begin{eqnarray}
\ddot{\chi}_{k}+3H\dot{\chi}_{k} +\Big[ \frac{k^{2}}{a^{2}} + m^{2}_{\chi} + \sigma\phi\Big]\chi_{k} = 0\,. \label{Potm2}
\end{eqnarray}
It is straightforward to show that the above equation can be also changed to the following standard Mathieu equation:
\begin{eqnarray}
\frac{d^{2}Y_{k}}{dz^{2}} + \left(A_{k} - 2q\cos(2z)\right)Y_{k} = 0\,, \label{Mathieu}
\end{eqnarray}
where
\begin{eqnarray}
A_{k} = \frac{4}{{\tilde M}^{2}}\left(k^{2} + m^{2}_{\chi}\right)\,,\,\, q = \frac{4\sigma\Phi}{\tilde{M}^{2}} \,, \label{Matheiu2}
\end{eqnarray}
where $\tilde{M}$ is defined in Eq.(\ref{Einf2}). Analogously, we can follow our analysis of the preceding subsection to obtain the general solution of the Mathieu equation.  In order to guarantee enough efficiency for the particle production, the Mathieu equation's parameters should satisfy the broad-resonance condition, i.e., $q\gg 1$. Therefore, in order to satisfy a broad resonance condition, we find that 
\begin{eqnarray}
g^{2}\gg \frac{10^{-5}}{\left((\alpha -1) \sinh ^{-1}\left(\frac{2}{\sqrt{3\alpha }}\right) (4 N-3 \alpha )^2\right)}\,. \label{qq}
\end{eqnarray}
For values of the parameters which work for our model, we then take $N=60,\,\alpha =4.00$, and find $g\gg 1.08\times 10^{-5}$. We also find the energy density of the field to be $\rho \sim {\tilde M}^{2}\phi^{2}\sim \frac{1}{25}{\tilde M}^{2} M^{2}_{p}$. This approximation allows us to further estimate the Hubble constant and we find that the Hubble rate would then be $H=\sqrt{\frac{1}{3 M^{2}_{p}}\rho}\sim {\tilde M}/(5\sqrt{3}) \sim 0.1 {\tilde M}$ which is the same as that of the preceding model. We can follow calculations given in the previous model. However, the results are very analogous to those found in the E-model. Therefore, we do not repeat it here and recommend readers the work of Ref.\cite{Kofman:1997yn} for much more detailed discussions. 

%%%%%%%%%%%%%%%%%%%%%%%
\section{Expansion of space}\label{ch4}
%%%%%%%%%%%%%%%%%%%%%%%
Unlike the preceding section, the background space-time curvature is worth investigating during preheating. As is known, the expansion of space causes the amplitude of inflaton oscillations to decay, while co-moving wave-numbers are red-shifted to smaller physical values. Let's take only the first model and consider our parametric resonance approach, we can see that the equation of motion for the scalar matter fields, Eq.(\ref{Pot2111}), can still be written in the form of a simple harmonic oscillator with a time varying frequency. Using the the same technique of performing Fourier transformation and taking $Y_{k}(t) = a(t)^{3/2}\chi_{k}(t)$, we have
\begin{eqnarray}
\ddot{Y}_{k} + \omega^{2}(k,t)\,Y_{k} = 0\,, \label{Pot2111e}
\end{eqnarray}
where a time dependent frequency of $Y_{k}(t)$ in this case is given by
\begin{eqnarray}
\omega^{2}(k,t) = \frac{k^{2}}{a^{2}} + m^{2}_{\chi}+ \sigma\Phi(t)\sin(M(t-t_{os}))-\frac{9}{4}H^{2}-\frac{3}{2}{\dot H}\,. \label{Pot20e}
\end{eqnarray}
It is clear that the above equation is not the Hill's equation anymore. However, we can quantitatively depict the effects of the space expansion on particle production by adding flow lines to the Floquet chart which can be trace the evolution of particular co-moving modes. Note that in the matter-dominated background which we are considering the last two terms on the right-hand side cancel. Notice that in the Einstein frame, our present model is equivalent to the trilinear one. Therefore, since ${\cal U}(\phi)\sim M^{2} \phi^{2}/2$, and $\Phi\propto a^{-3/2},\,k\propto a^{-1}$ and $3H^{2}=-2{\dot H}$, a given co-moving mode $k$ flows along $\Phi\sim k^{3/2}_{phys}\equiv (k/a)^{3/2}$ curve in $k_{phys}-\Phi$ plane, see Fig.\ref{plotpm}.

\begin{figure}[!h]	
	\includegraphics[width=12cm]{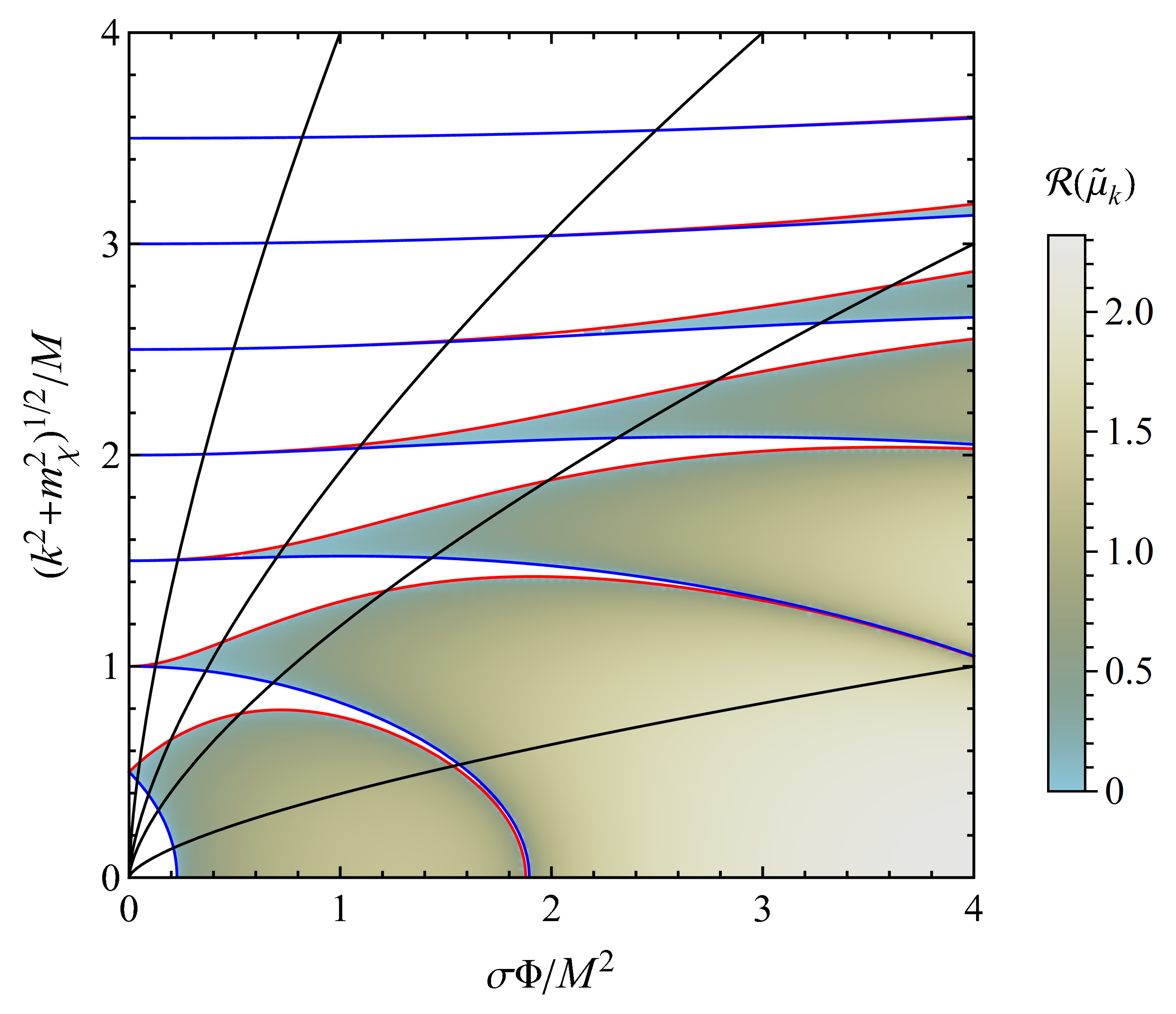}
	\centering
	\caption{We plot the instability chart featuring the real part of the Floquet exponent, characterizing the $\chi$ particle production rate in our model, ${\cal U}(\phi,\chi) \simeq M^{2}\phi^{2}/2 + m^{2}_{\chi}\chi^{2}/2 + \sigma \phi \chi^{2}$. The equation of motion for $\chi_{k}$ can be reduced to the Mathieu equation, Eq.(\ref{Matheiu2}), with $A_{k}= 4(k^{2}+ m^{2}_{\chi})^{1/2}/M,\,q=4\sigma \Phi/M^{2}$, where $\Phi$ is the amplitude of inflaton oscillations (see also Fig.\ref{plotp1}). In expanding universe $\Phi \propto a^{-3/2}$ and $k\propto a^{-1}$, implying that a given co-moving mode flows towards the bottom left corner of the chart as the universe expands as indicated with the black lines drawn for $m_{\chi }=0$ for simplicity.} \label{plotpm}
\end{figure}

The scale factor reads $a(t)\sim a_{0}(t/t_{0})^{2/3}$ and the amplitude of oscillations is decreasing as the universe expands, $\Phi\sim 0.98 M_{p}a^{-3/2}$. The parameter $q = \frac{4\sigma\Phi}{M^{2}}$ depends on time via $\Phi\propto a^{-3/2}$. Additionally, we can check in which a resonance band in our process develops. Following Ref.\cite{Kofman:1997yn}, the number of the band in the theory of the Mathieu equation is given by $n=\sqrt{A}$. In our case, reheating occurs for $A\sim 2q$, i.e. $n\sim \sqrt{2q}\sim \sqrt{8\sigma\Phi/M^{2}}$. Suppose we have an inflationary theory with $M\sim 10^{-6} M_{p}$, and let us take as an example $g\sim 10^{-1}$. Then after the first oscillation of the field, according to Eq.(\ref{Phi}), we have $\Phi(t)\sim M_{p}/6$, which corresponds to $q\sim 800$. This gives the band number $n \sim 40$. We close our discussion of the model by examining for very small $\phi(t)$ the change in the frequency of oscillations $\omega(t)$ which ceases to be adiabatic. The standard condition necessary for particle production is the absence of adiabaticity in the change of $\omega(t)$ \cite{Kofman:1997yn}:
\begin{eqnarray}
\frac{d\omega}{dt} \geq \omega^{2}\,.\label{adia}
\end{eqnarray}
For a narrow resonance, the above condition is not necessary since a small variation of
$\omega(t)$ may be exponentially accumulated in the course of
time. Nevertheless, it is not the case for a broad resonance, which implies that the condition (\ref{adia}) should be satisfied. Therefore our condition (\ref{adia}) implies that
\begin{eqnarray}
\frac{k^{2}}{a^{2}} \leq \Big(\frac{\sigma M\Phi}{2}\Big)^{2/3}-\sigma \phi\,,\label{adiaa}
\end{eqnarray}
where we have used that ${\dot \phi}\approx M\Phi$. Let us consider those momenta $k$ which satisfy the above condition as a function of $\phi(t)$. This condition becomes satisfied for small $k$ when the field $\phi(t)$ becomes smaller than $(M\Phi/2\sqrt{\sigma})^{2/3}$. Note that the expansion of space makes broad resonance more effective since more $k$ modes are red-shifted into the instability band as time proceeds.

%%%%%%%%%%%%%%%%%%%%%%%
\section{Discussion and outlook}
%%%%%%%%%%%%%%%%%%%%%%%
We have investigated the production of particles due to parametric resonances in model of inflation in general scalar-tensor theories
of gravity. We studied two models in which an inflaton field $\phi$ coupled to another scalar field $\chi$ with the interaction term $g^{2}\phi^{2}\chi^{2}$. We considered two-field scenario and then derived the potential of exponential and hyperbolic tangent forms. We studies the evolution of the background system when the back reaction on the background field is neglected. We examined the particle production due to parametric resonances in two models. We found that in Minkowski space the stage of parametric resonances can be described by the Mathieu equation. We demonstrated that parametric resonances in our models are sufficiently broad possible for the exponential growth of the number of particles.  

For a broad resonance to be satisfied, $q\gg 1$, we discover for E model that taking $N=60,\,\alpha\approx 1.333$, $g\gg 2.80\times 10^{-5}$, while for T model by using $N=60,\,\alpha =4.00$, we have $g\gg 1.08\times 10^{-5}$. We demonstrated that particle production in the two models is potentially efficient causing the number of particles $n_{k}$ in this process exponentially increases. More concretely, the number of $\chi$-particles grows as $\exp(qz)$, which in our case is equal
to $n_{k}\sim\exp(qMt/2)=\exp\big(\sqrt{3\alpha}g^{2}M^{3}_{p}Mt/(8\sqrt{2}\xi V_{0}\Phi)\big)$. Moreover, we also consider the case in which the expansion of space is taken into account. In this case, the expansion of space makes broad resonance more effective since more $k$ modes are red-shifted into the instability band as time proceedsProgress.

However, in the second stage of preheating, backreaction might increase the frequency of oscillations of the inflaton field, which makes the process even more efficient. Another issue is that a model with the quantum scalar field $\chi$ non-minimally coupled to gravity is another interesting scenario. Moreover, reheating mechanism of inflation in general scalar-tensor theories of gravity is worth investigating. We recommend readers to Ref.\cite{Kofman:1997yn} for detailed discussion on the topics. However, we will leave these interesting topics for our future investigation.

\section*{Acknowledgements}
P. Channuie acknowledged the Mid-Career Research Grant 2020 from National Research Council of Thailand (NRCT5-RSA63019-03) and is partially supported by the National Science, Research and Innovation Fund (SRF) with grant No.R2565B030.

\end{document}